\documentclass[prb,twocolumn,aps,showpacs,amsmath]{revtex4}
\usepackage{graphicx,bm,amsmath,amssymb,psfrag}
\newcommand{\mscr}[1]{{\mbox{\scriptsize #1}}}
\begin{document}
\title{Spin susceptibilities, spin densities and their connection to
  spin-currents} 
\author{Sigurdur I.\ Erlingsson} 
\author{John Schliemann} 
\author{Daniel Loss} 
\affiliation{Department of Physics
  and Astronomy, University of Basel, Klingelbergstrasse 82, CH-4056,
  Switzerland}
\begin{abstract}
We calculate the frequency dependent spin susceptibilities for a
two-dimensional electron 
gas with both Rashba and Dresselhaus spin-orbit interaction.  
The resonances of the susceptibilities depend on the relative values
of the Rashba and   
Dresselhaus spin-orbit constants, which could be manipulated by gate voltages.
We derive exact continuity
equations, with source terms, for the spin density and use those to
connect the spin current to the spin density.  
In the free electron
model the susceptibilities play a central role
in the spin dynamics
since both the spin density and the spin current are
proportional to them.
\end{abstract}
\pacs{73.63.-b,85.75.-d} 
\maketitle
\section{Introduction}
The ability to manipulate spin states in semiconducting, and
metallic, nanostructures is now the focus of much experimental and
theoretical attention.  Some spintronics application are
already in use\cite{wolf01:1488,prinz98:1660} and more have been proposed,
ranging from spin
FET's\cite{datta90:665,schliemann03:146801,egues03:2658} to spin 
qubits.\cite{loss98:120,awschalom02:xx} 
In many cases the spin-orbit interaction can be used to manipulate the
electron spin via electronic means.\cite{kato04:50} This
sort of electronic manipulation is important since technologically,
electric field control of spins is preferred over magnetic field
control. Also, electronic control of spins has revealed many interesting
physics in experiment involving semiconductor
heterostructures.\cite{kikkawa99:139,salis01:619} 

Already quite some time ago, it was proposed that electric fields
could lead to magnetization in anti-ferromagnetic materials fulfilling
certain symmetries\cite{dzyaloshinkii59:628}, this is the so called
magnetoelectric 
effect.   The magnetoelectric effect was
considered for conductors with special symmetry
properties\cite{levitov85:133} and for pyro-electric 
superconductors\cite{edelstein95:2004}.
Electric field induced spin-orientation in semiconductors due to linear in
momentum spin-orbit interaction was also discussed in Refs.\
\onlinecite{aronov89:431,edelstein90:233,magarill00:195}.
There has been renewed interest in this subject since this
induced spin polarization might serve as spin-injectors in certain
semiconductor 
heterostructures.\cite{voskoboynikov00:387,cartoixa01:309}
Only recently there have been experiments which seem to demonstrate
such current induced polarization of spins via spin-orbit
interaction.\cite{kato:0403407,ganichev:0403641}

A somewhat related effect is the so called spin-Hall effect.  In the 
normal Hall effect the electrons are deflected by the Lorentz force,
but in the spin-Hall effect spins are scattered by impurities
preferentially to the 
left (right) if their spins point 'up' 
('down') due to spin-orbit interaction.\cite{hurd73:01,chien80:01}
This will result in   
a spin current, but no net charge current, flowing perpendicular to
the applied charge current. 
Such an extrinsic transverse spin current was already investigated by
Dyakononov and Perel
some time ago in Ref.\ \onlinecite{dyakonov71:467} and more recently
by Hirsch\cite{hirsch99:1834}.  
Presently there is much interest in
spin-orbit 
mediated spin-Hall effect in semiconducting heterostructures. 
Here
the effect can appear in
hole\cite{murakami03:1348,culcer:0309475,hu04:5991,bernevig:0311024,hu:0401231,murakami04:241202R,schliemann:0405436}
or
electron\cite{sinova04:126603,schliemann04:165315,sinitsyn:0310315,shen:0310368,burkov:0311328,rashba03:241315R,inoue:0402442,shen:0403005,xiong:0403083,rashba:0404723,dimitrova:0405339}
doped semiconductors
due to band structure properties and
impurities are not necessary.
Thus, the term intrinsic spin Hall effect
is used to distinguish this from the previously discussed mechanism
since it occurs already in the absence of
impurities.\cite{sinova04:126603,culcer:0309475}
Due to differences in the band structure of the holes and electron
there is an important distinction between the two since in the electron system
the spin Hall conductance (the ratio of the transverse spin current
and the applied electric field) reaches a universal value of $e/8\pi$
in clean systems.\cite{sinova04:126603}
However, impurities are believed to modify this universal
value\cite{schliemann04:165315,xiong:0403083} 
and in addition,  the  exact influence of vertex
corrections on the spin-Hall conductivity is currently under
investigation\cite{inoue:0402442,murakami:0405003,dimitrova:0405339}. 

In this paper we consider the spin susceptibilities of a
2DEG with Rashba and Dresselhaus spin-orbit coupling.  Due to the
strong 2DEG confinement the Dresselhaus coupling reduces to terms
linear in momentum.  For such linear momentum spin-orbit coupling the spin
susceptibility can be used to characterize other transport
properties.  
We calculate the susceptibilities using a free electron model and relate
them to the electric field induced spin density.  
Also, we derive continuity equations (with source terms)
for the spin density and spin current, similar to the equations
already derived for only Rashba coupling.\cite{burkov:0311328}  These
equations are exact operator identities and via them we can relate the
spin current to the spin density. 
Via these relations the  spin current (which is non-trivial to measure)
can be connected to the spin density, or magnetization, which is
easier to detect. 
The susceptibilites play a central role in the free electron model,
since the Fourier transform of the spin density and the spin current
are proportional to the susceptibilities.
\section{The Model}
We consider a two dimensional electron gas (2DEG) with Rashba and
Dresselhaus spin-orbit coupling.  In the absence of external fields
the Hamiltonian may be written as  
\begin{equation}
H=\frac{p_x^2+p_y^2}{2m}+
\frac{\alpha}{\hbar}(p_y{\sigma}_x-p_x{\sigma}_y)
+\frac{\beta}{\hbar}(p_y{\sigma}_y-p_x{\sigma}_x),
\label{eq:hamiltonian}
\end{equation}
where $\alpha$ and $\beta$ are the Rashba and Dresselhaus (linear)
coefficient in a 2DEG.
It is easy to see that $[p_x,H]=[p_y,H]=0$, and thus we seek
eigenstates of the form
\begin{equation}
\psi_{\bm{k},s}(\bm{r})\equiv \langle \bm{r}|\bm{k}s
\rangle=\frac{e^{i\bm{k}\cdot\bm{r}}}{\sqrt{A}} 
\bm{u}_s(\bm{k})
\end{equation}
where $\bm{u}_s(\bm{k})$ is spinor to be determined and $A$ is the
system area.  Since the momenta are conserved it is possible to introduce an
effective magnetic field
\begin{equation}
\bm{\Gamma}=\left (
\begin{array}{c}
-\beta k_x+\alpha k_y\\
\beta k_y-\alpha k_x\\
0
\end{array}
\right ).
\end{equation}
The eigenspectrum can be written in terms of this effective magnetic
field, the eigenenergies of the Hamiltonian being ($s=\pm1$)
\begin{equation}
E_s(\bm{k})=\frac{\hbar^2k^2}{2m}+s|\Gamma^+|,
\label{eq:eigenvalues0}
\end{equation}
and the corresponding spinor 
\begin{eqnarray}
\bm{u}_s(\bm{k})&=&\frac{1}{\sqrt{2}}
\left (
\begin{array}{c}
1\\
s\frac{\Gamma^+}{|\Gamma^+|}
\end{array}
\right ),
\label{eq:eigenspinors0}
\end{eqnarray}
where $\Gamma^+=\Gamma_x+i\Gamma_y$.  
Note that Eqs.\ (\ref{eq:eigenvalues0}) and (\ref{eq:eigenspinors0})
are also valid for an  in-plane magnetic field 
$\bm{B}$ using the substitution  $\bm{\Gamma} \rightarrow
\bm{\Gamma}+\frac{1}{2}g\mu_B \bm{B}$, where $g$ and $\mu_B$ are the
effective $g$-factor and Bohr magneton, respectively.

Writing the quasi-momentum in polar coordinates
$\bm{k}=k(\cos\theta, \sin\theta)$, 
the eigenspectrum of the Hamiltonian in Eq.\ ({\ref{eq:hamiltonian}}) 
becomes
\begin{equation}
E_s(\bm{k})=\frac{\hbar^2 k^2}{2m}+s k\sqrt{(\alpha^2+\beta^2)
g(\theta)},
\label{eq:eigenenergies}
\end{equation}
where $g(\theta)=1-\sin(2\phi) \sin(2\theta)$ determines the
anisotropy of the Fermi surfaces and the corresponding eigenspinors
are 
\begin{eqnarray}
\bm{u}_s(\bm{k})&=&\frac{1}{\sqrt{2}}
\left (
\begin{array}{c}
1\\
-s\frac{(\cos(\phi)e^{-i\theta}+i\sin(\phi)e^{i\theta})}{g(\theta)}
\end{array}
\right ).
\label{eq:eigenvectors}
\end{eqnarray}
Here we have introduced the following
parameterization of the spin-orbit coupling strength
\begin{equation}
\sin(\phi)=\frac{\alpha}{\sqrt{\alpha^2+\beta^2}},\quad
\phi \in[-\pi/2,\pi/2].
\end{equation}
The eigenfunctions in Eq.\ (\ref{eq:eigenvectors}) have the
interesting property that they 
depend on the spin-orbit coupling parameters $\alpha$ and $\beta$ only
via the angle $\phi$.
This allows one to take the limit $\alpha,\beta \rightarrow 0$ such
that the angle $\phi$ remains fixed and the resulting eigenvectors
in Eq.\ (\ref{eq:eigenvectors}) are also (degenerate) eigenvectors
of the  free electron Hamiltonian.\cite{molenkamp01:121202R}  As was
pointed out in Ref.\ \onlinecite{rashba03:241315R}, the Kramers
conjugate state of $\psi_{\bm{k},s}(\bm{r})$ is
$\psi_{-\bm{k},s}(\bm{r})$, i.e.\ they belong to the same branch. 
\section{Spin susceptibilities and current induced magnetization}
For a weak driving field the response of the system is obtained by
the Kubo formalism.  Due to the spin-orbit coupling, a pure electric
field driving results in a non-zero magnetic response.
Since the spin-orbit term in Eq.\ (\ref{eq:hamiltonian}) is linear in momenta,
both the response functions due to magnetic and electric perturbation
can be expressed with spin susceptibilities. 
The $\eta=x,y,z$ component of the spin density operator is defined as 
\begin{equation}
\rho_\eta(\bm{r})
=\sum_n \sigma_{n,\eta}(\bm{r})=
\sum_n \sigma_{n,\eta} \delta(\bm{r}-\bm{r}_n),
\label{eq:spinDensityOperator}
\end{equation}
where $\bm{r}_n$ and $\sigma_{n,\eta}$ are the position operator and
Pauli matrix, respectively, of the $n$th electron.  
For a translationally invariant system, the wave
vector and frequency ($\bm{q},\omega$) dependent susceptibilities are   
\begin{eqnarray}
\chi_{\eta \eta'}(\bm{q},\omega)
\!&\!=\!&\!
\int_0^\infty \!\! dt e^{i\omega t}
\frac{i}{\hbar A}
\langle [{\rho}_{\eta}(\bm{q},t),{\rho}_{\eta'}(-\bm{q})]\rangle,
\\
\!&\!=\!&\!
\int_0^\infty \!\! dt e^{i\omega t}
\frac{1}{A}\sum_{\bm{k},s} 
\frac{i f_{\bm{k}s}}{\hbar}
\langle
[{\sigma}_{\eta}(\bm{q},t),{\sigma}_{\eta'}(-\bm{q})]\rangle_{\bm{k}s},
\nonumber \\
\label{eq:susceptibilities}
\end{eqnarray} 
where $f_{\bm{k}s}=f(E_s(\bm{k}))$ with $f$ being the Fermi
distribution function and we used the notation
$\langle \dots \rangle_{\bm{k}s}=\langle \bm{k}s|\dots |\bm{k}s
\rangle$.
The frequency should be viewed   
as $i \omega \rightarrow i(\omega+i\hbar^{-1}\epsilon)$ to regularize the
integral. 
The operators in Eq.\ (\ref{eq:susceptibilities}) refer to
single particle operators.
This susceptibility is a spin density response function and to get the
magnetization response function, each 
spin density operator 
in Eq.\ (\ref{eq:susceptibilities}) should be multiplied with the
electron effective magnetic moment $g\mu_B/2$.  
Using Eq.\ (\ref{eq:susceptibilities}) and the eigenspectrum
represented by Eqs.\ (\ref{eq:eigenenergies}) and  (\ref{eq:eigenvectors})
the susceptibilities for a spatially
homogeneous perturbation $(\bm{q}=0)$ become
\begin{widetext} 
\begin{eqnarray} 
\chi_{xx}(\omega)&=&
\frac{1}{(2\pi)^2\hbar}\int_0^{2\pi}d\theta\frac{(\alpha\sin\theta-\beta
\cos\theta)^2}{(\alpha^2+\beta^2)^{1/2}\sqrt{g(\theta)}}
\int_{k_+(\theta)}^{k_-(\theta)}
dk
\frac{k^2}{4(\alpha^2+\beta^2)g(\theta)k^2-(\hbar \omega+i
  \epsilon)^2} 
\label{eq:definition_chi_xx} \\
\chi_{xy}(\omega)&=&
\frac{1}{(2\pi)^2\hbar}\int_0^{2\pi}d\theta\frac{(\alpha\sin\theta-\beta
\cos\theta)(\beta \sin \theta -\alpha \cos
\theta)}{(\alpha^2+\beta^2)^{1/2}\sqrt{g(\theta)}} 
\int_{k_+(\theta)}^{k_-(\theta)}
dk
\frac{k^2}{4(\alpha^2+\beta^2)g(\theta)k^2-(\hbar \omega+i
  \epsilon)^2} .
\label{eq:definition_chi_xy}
\end{eqnarray}
\end{widetext}
Note that all $\bm{q}=0$ dependence has been dropped for clarity. 
Here we assumed zero temperature and the Fermi
distribution function was replaced by a step function.  The
resulting Fermi contours $k_\pm(\theta)$ are the solutions of 
\begin{equation}
k_{F}^2=k_\pm(\theta)^2 \pm k_\pm(\theta)k_{SO}\sqrt{g(\theta)},
\end{equation}
where $k_F^2=2m\varepsilon_F/\hbar^2$ is the squared Fermi momentum and
$k_{SO}=m\sqrt{\alpha^2+\beta^2}/\hbar^2$.
The $k$ integral results in a linear term and 
a term involving an inverse tangent
in $k_\pm(\theta)$ and the subsequent angular integrals cannot be
solved analytically.  The inverse tangent can be 
expanded in  powers of $k_{SO}/k_{F}\ll 1$, resulting in
the following lowest order result
\begin{equation} 
\chi_{xx}(\omega)=\chi_{yy}(\omega)=
\frac{m}{2\pi \hbar^2} 
\Bigl (1+ 
\frac{(\hbar\omega+i \epsilon)^2}
{\prod_{s}\sqrt{\varepsilon_s^2-(\hbar \omega +i\epsilon)^2}}
\Bigr ),
\label{eq:chi_xx}
\end{equation}
where the resonance energies are
$\varepsilon_\pm^2=8\varepsilon_{SO}\varepsilon_F(1\pm \sin(2
\phi))$, with $\varepsilon_{SO}=m(\alpha^2+\beta^2)/\hbar^2$.
Using the same procedure we can 
calculate the off-diagonal susceptibilities in a similar manner
\begin{equation}
\chi_{xy}(\omega)=\chi_{yx}(\omega)=\frac{\delta\chi(\omega)-
  \chi_{xx}(\omega)}{\sin(2\phi)} 
-\sin(2\phi)\delta\chi(\omega),
\label{eq:chi_xy}
\end{equation}
where we have defined
\begin{eqnarray}
\delta\chi(\omega)&=&\frac{m}{2\pi \hbar^2}
\frac{8\varepsilon_{SO}\varepsilon_F}
{\prod_{s}\sqrt{\varepsilon_s^2-(\hbar \omega +i\epsilon)^2}}.
\label{eq:delta_chi}
\end{eqnarray}
The magnetization is related to the spin density in Eq.\
(\ref{eq:spinDensityOperator}) through $\bm{m}(\bm{r})=\frac{1}{2}g
\mu_B \bm{\rho}(\bm{r})$ which leads to the standard linear response
relation 
\begin{equation}
m_\eta(\bm{q},\omega)=\Bigl (\frac{1}{2}g\mu_B \Bigr )^2
\chi_{\eta\eta'}({\bm{q},\omega})B_{\eta'}({\bm{q},\omega}) .
\label{eq:magnetization}
\end{equation}
To obtain the Pauli paramagnetic susceptibility one should take the
following order of limits\cite{mahan00:xx}
\begin{equation}
\lim_{\bm{q}\rightarrow 0}
\lim_{\omega \rightarrow 0}
\chi_{\eta\eta'}({\bm{q},\omega})
=\frac{m}{\pi \hbar^2}\delta_{\eta,\eta'}.
\label{eq:pauli}
\end{equation}
The diamagnetic contribution can be disregarded since we assume an in-plane
magnetic field. 
The susceptibilities in Eqs. (\ref{eq:definition_chi_xx}) and
(\ref{eq:definition_chi_xy}) are calculated for the reverse order 
of limits done in Eq.\ (\ref{eq:pauli}).  
These $\omega 
\neq 0$ susceptibilities 
are the spin-orbit
contribution
coming from the region in $\bm{k}$-space where
only one ($s$=$-1$) of the two branches is occupied.\footnote{Assuming
  that $v_F q\ll k_BT \ll
  \sqrt{\varepsilon_\mscr{SO}\varepsilon_F}$, the dominant  
  contribution comes 
  from the region between the two  Fermi contours (the
  spin-orbit contribution) if $\hbar \omega \gg \varepsilon_F
  q/k_F$.}  
Also, the spin current (see discussion below) which results from the
spin-orbit interaction is non-zero due to contributions from
the $\bm{k}$-space area between the two Fermi
contours.\cite{sinova04:126603,schliemann04:165315,rashba:0404723}
Thus we only focus on this contribution when we relate the
susceptibilities to the spin-orbit mediated spin densities and the spin
currents. 

The value of the resonance  frequency is determined by 
$\sqrt{8\varepsilon_{SO}\varepsilon_F} \approx 0.16$\,meV=$40$\,GHz
for typical GaAs parameters\cite{miller03:76807}: $\alpha=0.5 
\times10^{-9}$\ meV\,m and electron density $n_e=4 \times
10^{15}$\,m$^{-2}$.   For lower frequencies  the susceptibilities
remain nominally constant.   
In the limit $\alpha \ll \beta$ the lowest order contribution to
Eqs.\ (\ref{eq:chi_xx}) and (\ref{eq:chi_xy}) become
\begin{eqnarray} 
\chi_{xx}(\omega)&=& 
\frac{m}{2\pi \hbar^2}  
\frac{1}{1-\frac{(\hbar \omega +i
    \epsilon)^2}{8\varepsilon_{SO}\varepsilon_F}}\\
\chi_{xy}(\omega)&=&-\frac{m}{2\pi \hbar^2}
\frac{\alpha}{\beta}
\frac{1-\frac{(\hbar \omega +i
    \epsilon)^2}{4\varepsilon_{SO}\varepsilon_F}}{\left (1-\frac{(\hbar \omega +i
    \epsilon)^2}{8\varepsilon_{SO}\varepsilon_F}\right )^2}.
\end{eqnarray} 
Here we have not included impurities and thus the regularization 
parameter $\epsilon$ can strictly only be attributed to an 
adiabatic turning on of the external electric or magnetic field.   

In the absence of electric and magnetic fields, the spin-orbit 
interaction does not give rise to a net magnetization.  Even though 
the spin-orbit interaction has the form of a momentum-dependent magnetic
field, the total contribution averages to 
zero.\cite{winkler04:45317,winkler:0401067}
However, although the spin-orbit induced splitting does not give
rise to an equilibrium magnetization, there is an asymmetry in the local
magnetic field in momentum space, i.e.\ the local magnetic field is odd
under wave vector reversal, and any translation of the
Fermi sphere away from the $\Gamma$-point will induce a
magnetization.\cite{edelstein90:233} 
Applying a homogeneous electric field
$\bm{E}(\bm{r},t)=\bm{E}^0e^{-i\omega t}$ to the system will give rise
to the following time dependent perturbation 
\begin{equation}
V(t)=
-\frac{e}{i\omega} \bm{E}^0 e^{-i\omega t}\cdot {\bm{j}}(\bm{q}=0) 
\end{equation}  
where ${\bm{j}}(\bm{q})$ is the Fourier transform of 
the current density operator 
\begin{eqnarray} 
{\bm{j}}(\bm{r})&=&\sum_n {\bm{j}}_n (\bm{r})=\sum_n \frac{1}{2}\{ 
\delta(\bm{r}-{\bm{r}}_n)  
,{\bm{v}}_n \} , 
\label{eq:currentDensityOperator} 
\end{eqnarray} 
and the velocity operator ${\bm{v}}_n$ for the Hamiltonian in Eq.\ 
(\ref{eq:hamiltonian}) is given by 
\begin{eqnarray} 
{\bm{v}}_n&=& 
\left ( 
\begin{array}{c} 
\frac{1}{m}p_{n,x}-\frac{\alpha}{\hbar}{\sigma}_{n,y}-
\frac{\beta}{\hbar}{\sigma}_{n,x}
\\
\frac{1}{m}p_{n,y}+\frac{\alpha}{\hbar}{\sigma}_{n,x}+
\frac{\beta}{\hbar}{\sigma}_{n,y}
\end{array} 
\right ). 
\label{eq:velocityOperator} 
\end{eqnarray} 
From linear response theory the Fourier transform of the electric 
field induced spin density may be written as 
\begin{eqnarray} 
\langle{\rho}_x(\omega)\rangle&=& 
\frac{eE_x(\omega)}{i\omega} \Bigl ( 
\frac{\alpha}{\hbar}\chi_{x y}(\omega)+\frac{\beta}{\hbar} 
\chi_{x x}(\omega) \Bigr ) 
\nonumber \\ 
& &-\frac{eE_y(\omega)}{i\omega} \Bigl ( 
\frac{\alpha}{\hbar}\chi_{x x}(\omega)+\frac{\beta}{\hbar} 
\chi_{x y}(\omega) \Bigr ) \label{eq:spinDensity_x}  \\
\langle{\rho}_y(\omega)\rangle&=& 
\frac{eE_x(\omega)}{i\omega} \Bigl ( 
\frac{\alpha}{\hbar}\chi_{xx}(\omega)+\frac{\beta}{\hbar} 
\chi_{xy}(\omega) \Bigr ) 
\nonumber \\ 
& &-\frac{eE_y(\omega)}{i\omega} \Bigl ( 
\frac{\alpha}{\hbar}\chi_{xy}(\omega)+\frac{\beta}{\hbar} 
\chi_{xx}(\omega) \Bigr ).
\label{eq:spinDensity_y} 
\end{eqnarray} 
This result is reminiscent of the pure magnetic field induced 
spin density, except here $eE_y(\omega)/i\omega$ plays the role of  
magnetic field, via the spin-orbit coupling.   
As was pointed out in Refs.\ 
\onlinecite{rashba:0404723,edelstein90:233} the dc limit corresponds  
to replacing the frequency with momentum scattering $i\omega
\rightarrow -1/\tau$. 

Multiplying Eqs.\ (\ref{eq:spinDensity_x}) and
(\ref{eq:spinDensity_y}) by the sample area will give the total  
number of induced magnetic moments, measured in units of $g\mu_B/2$.  
Applying an electric field $E\approx 100$\,V/cm to a
GaAs 2DEGs with high mobility ($\hbar/\tau\approx10^{-2}$\,meV)
and a sample area of $A=(500\,\mu$m$)^2$, the number of magnetic
moments (Bohr magnetons) would be around $2.5\times10^7$.  For a
2DEG thickness of a few nm these magnetic moments produce
a magnetic field of the order $10^{-6}$\,T.
Probing 2DEG properties using ESR techniques has been succesfully used
to determine the spin-orbit splitting\cite{stein83:130} and other 2DEG spin
properties\cite{dobers88:5453,nestle97:4359R}.  Using similar ESR  
techniques, the spin-orbit coefficients $\alpha,\beta$ could in
principle be determined by 
measuring different spin density component for different direction of 
driving current, as a function of $\alpha$ which could be tuned by
gate voltages.\cite{grundler00:6074,koga02:46801,miller03:76807} Such
an induced spin density could be detected by Faraday
rotation measurements.\cite{kato:0403407}
\section{Connection to spin current} 
The usual way of deriving the operator version of the particle 
continuity equation is to start from the definition of the 
density operator
\begin{eqnarray} 
{\rho}(\bm{r},t)=\sum_n \delta(\bm{r}-{\bm{r}}_n(t)),
\end{eqnarray}
and from there one can derive the standard continuity equation 
\begin{equation} 
\frac{\partial }{\partial t}{\rho}(\bm{r},t)+\nabla \cdot {\bm{j}}(\bm{r},t)=0,
\label{eq:particleContinuity}
\end{equation}
from the Heisenberg equation of motion for ${\rho}(\bm{r},t)$.  The form
of the current density in Eq.\ (\ref{eq:particleContinuity}) is
uniquely determined by the Heisenberg equation of motion.  For the
Hamiltonian in Eq.\ (\ref{eq:hamiltonian}) the current density is
given by Eq.\ (\ref{eq:currentDensityOperator}), using the velocity operator
in Eq.\ (\ref{eq:velocityOperator}).  
In the case of the spin density operator
\begin{equation}
\rho_\eta(\bm{r},t)=
\sum_n
\sigma_{n,\eta}(t)\delta(\bm{r}-{\bm{r}}_n(t)),
\end{equation}
the same procedure of evaluating the Heisenberg equation motion will
not result in a unique definition of the associated spin current
density.  The reason is that the  precession due to the momentum
dependent spin-orbit magnetic field introduces additional terms into the
equation of motion.  In order to proceed one has to postulate a
form for the spin current density.  The most widely used definition,
and physically reasonable, is the following  
\begin{equation}
{\bm{j}}^{\eta}(\bm{r})=\sum_n\frac{1}{2}\{
{\sigma}_{n,\eta},{\bm{j}}_n(\bm{r}) \}.
\label{eq:spinCurrentDensityOperator}
\end{equation}
This form of the spin current is Hermitian and reduces to the 
standard spin current form when the velocity operator is 
spin independent.\cite{brataas01:99} 
Having determined the form of the spin current the resulting 
continuity equations become   
\begin{eqnarray}  
\frac{\partial}{\partial t} {\rho}_x (\bm{r},t)
+\nabla\cdot{\bm{j}}^{x} (\bm{r},t)
\!&\!=\!&\! -\frac{2m\alpha}{\hbar^{2}}{j}^{z}_{x}(\bm{r},t)+
 \frac{2m\beta}{\hbar^{2}}{j}^{z}_{y} (\bm{r},t) \nonumber\\
\label{eq:spincont1}\\ 
\frac{\partial}{\partial t} {\rho}_y  (\bm{r},t)
+\nabla\cdot{\bm{j}}^{y}  (\bm{r},t)
\!&\!=\!&\! -\frac{2m\alpha}{\hbar^{2}}{j}^{z}_{y} (\bm{r},t)+
 \frac{2m\beta}{\hbar^{2}}{j}^{z}_{x}  (\bm{r},t)\nonumber \\
\label{eq:spincont2}\\ 
\frac{\partial}{\partial t} {\rho}_z  (\bm{r},t)
+\nabla\cdot{\bm{j}}^{z}  (\bm{r},t)
\!&\!=\!&\!+\frac{2m\alpha}{\hbar^{2}}({j}^{x}_{y} (\bm{r},t)+{j}^y_x
 (\bm{r},t) )\nonumber \\ 
 & &-\frac{2m\beta}{\hbar^{2}}({j}^{x}_{x} (\bm{r},t)+{j}_y^y
 (\bm{r},t)). \nonumber \\
\label{eq:spincont3} 
\end{eqnarray} 
Similar equations have already been derived for pure Rashba 
coupling.\cite{burkov:0311328}
The above Eqs.\ (\ref{eq:spincont1})-(\ref{eq:spincont3}) are exact
relation for a systems with spin-orbit coupling linear in momentum and
including impurities would not change their form.\footnote{ 
When Hamiltonians representing non-magnetic impurites 
$H_\mscr{imp}=H_\mscr{imp}({\bm{r}})$ are added to Eq.\ 
(\ref{eq:hamiltonian}), the commutator in the  
Heisenberg equation of motion for an operator 
${\mathcal{O}}({\bm{r}},{\bm{\sigma}})$ becomes  
$[{\mathcal{O}}({\bm{r}},{\bm{\sigma}}),H+H_\mscr{imp}]= 
[{\mathcal{O}}({\bm{r}},{\bm{\sigma}}),H]$,
where ${\mathcal{O}}$ is any operator that does not explicitly
depend on $\bm{p}$.  Both the particle and spin density are such 
operator and thus the equations of motion remain the same in the 
presence of impurities. The complications due to the  impurities 
appear in the calculation of the average values of the operators.
} 
\footnote{
Homogeneous ($\bm{q}$=0) spin systems driven by a constant field
$E_0$  
will exhibit damping of magnetization, i.e.\
$\partial_t \langle \rho_\eta(t\rightarrow \infty)\rangle=0$
for any finite amount of dissipation.
Then, from Eq.\ (\ref{eq:spincont1}) and the relation 
$ 
\lim_{t \rightarrow \infty}\partial_t  \langle \rho_x(t) \rangle =\lim_{s
  \rightarrow 0} s^2 \langle \tilde{\rho}_x(s) \rangle
$ 
for the Laplace transform, we get
$0=\lim_{s\rightarrow 0}s^2 \tilde{\rho}_x(s) 
=\frac{2m\beta E_0}{\hbar^2}\tilde{\sigma}_{xy}^z( s\rightarrow 0)
$,
where we used $\langle \tilde{j}_x^z(s)\rangle
=\frac{\hbar}{2}\tilde{\sigma}_{xy}^z(s) E_0/s$ and $\alpha=0$.
Thus,
the {\em dc} spin conductivity vanishes
for any 
dissipation mechanism.
The same follows directly from the other magnetization components and in the
case of $\alpha \neq 0$.}

Taking the thermal average of Eqs.\
(\ref{eq:spincont1})-(\ref{eq:spincont3}) gives partial differential
equation connecting the spin densities and spin currents.  
Based on these equations one can make a few observation on the
nature of the spin current, without explicitly solving them. 
First of all, for a homogeneous system in the stationary limit the rhs
of Eq.\ (\ref{eq:spincont3}) must vanish for all values of $\alpha$,
$\beta$.  This is trivially satisfied for all $\langle
\bm{j}^\eta\rangle =0$, but
more interestingly also when $\langle j_y^x\rangle =-\langle j_x^y
\rangle$ and $\langle j_x^x\rangle =-\langle j_y^y
\rangle$. It is easy to show that the latter case is true in
equilibrium 
\begin{eqnarray}
\langle j^{x}_{x}\rangle=-\langle j^{y}_{y}\rangle & = & 
\frac{1}{3\pi}\frac{\beta}{\hbar}\left(\frac{m}{\hbar^{2}}\right)^{2}
\left(\alpha^{2}-\beta^{2}\right) \\
\langle j^{y}_{x}\rangle=-\langle j^{x}_{y}\rangle & = & 
\frac{1}{3\pi}\frac{\alpha}{\hbar}\left(\frac{m}{\hbar^{2}}\right)^{2}
\left(\alpha^{2}-\beta^{2}\right),
\end{eqnarray}
which covers the results of Ref.\ \onlinecite{rashba03:241315R} as special cases.
In particular, the current expectation values vanish at $\alpha=\pm\beta$
due to the additional concerved quantity arising at these points
\cite{schliemann03:146801}.
Furthermore, Eq.\
(\ref{eq:spincont3}) shows these equilibrium currents do not act as
source terms for the spin density, since the rhs always vanishes.

Let us now consider a homogeneous system such that the divergence
terms vanish.
By using Eqs.\ (\ref{eq:spincont1}) and (\ref{eq:spincont2}) one can
derive the following
identity for the Fourier transform of
the  $x$ and $y$ component of the average $z$-polarized spin current
\begin{eqnarray} 
\langle {j}_x^z(\omega) \rangle= 
\frac{\hbar^2 i\omega (\alpha \langle {\rho}_x(\omega) \rangle
  +\beta 
\langle {\rho}_y(\omega) \rangle )}{2m(\beta^2-\alpha^2)}
\label{eq:spinCurrent_x_Exact} \\
\langle {j}_y^z(\omega) \rangle= 
\frac{\hbar^2 i\omega (\beta \langle {\rho}_x(\omega) \rangle
  +\alpha
\langle {\rho}_y(\omega) \rangle )}{2m(\beta^2-\alpha^2)}.
\label{eq:spinCurrent_y_Exact} 
\end{eqnarray}
These relations establish a connection between the spin 
current components $\langle {j}_{x,y}^z\rangle $ and the $x,y$
components of the spin  
density, in the frequency domain. 
This is quite useful since the spin current, which is hard to 
detect, is related  to a quantity which is easier to measure.  
Also, Eqs.\ (\ref{eq:spinCurrent_x_Exact}) and
(\ref{eq:spinCurrent_y_Exact}) is a good starting point for connecting
spin current and spin density response functions using standard Kubo
formalism. 

Let us now assume a homogeneous electric field applied in the $y$
direction. 
The spin conductivities are defined as the ratio of
the spin current and applied electric field
\begin{equation}
\sigma_{\eta y}^z(\omega)=\frac{\frac{\hbar}{2}\langle 
  {j}_\eta^z(\omega) \rangle}{E_y(\omega)},
\end{equation}
where the factor $\hbar/2$ in the definition of the spin
conductivity is due to our definition of the spin current in terms
Pauli matrices and not the spin operators, i.e.\
${\bm{S}}=\hbar\bm{\sigma}/2$.  
Using Eqs.\ (\ref{eq:spinDensity_x}) and
(\ref{eq:spinDensity_y}) to relate the spin density to the
susceptibilities we obtain the following result for the ac spin 
conductivities 
\begin{eqnarray} 
\sigma_{xy}^z(\omega)
&=&
\frac{e\hbar^2}{4m}
\frac{\alpha^2-\beta^2}{\beta^2+\alpha^2}
\delta \chi(w) \label{eq:spinCurrent_susceptibility_x} \\
\sigma_{yy}^z(\omega)
&=& 
-\frac{\alpha^2-\beta^2}{2\beta \alpha}
\frac{e\hbar^2}{4m}
\bigl (\delta \chi(w)-\chi_{xx}(w) \bigr ).
\label{eq:spinCurrent_susceptibility_y} 
\end{eqnarray} 
These equations for the spin conductivities, along with Eqs.\
(\ref{eq:chi_xx}) and (\ref{eq:delta_chi}),
in the $\omega \rightarrow 0$ limit reproduce the result in Ref.\
\onlinecite{sinitsyn:0310315} up to a sign convention for $\beta$.
For either pure Rashba or Dresselhaus
we have the following result 
\begin{eqnarray} 
\sigma_{xy}^z(\omega)&=&\pm\frac{e}{8\pi} 
\left ( 
\frac{1}{1- 
\frac{(\hbar \omega +i \epsilon)^2}{8\varepsilon_{SO}\varepsilon_F}} 
\right) \\
\sigma_{yy}^z(\omega)
&=& 0
\end{eqnarray} 
the upper (lower) sign refers to a pure Rashba 
(Dresselhaus).  
Here one can replace $\epsilon$ with the impurity scattering 
broadening  
$\hbar/\tau$, which reproduces the zero frequency result in Ref.\ 
\onlinecite{schliemann04:165315}. 
Taking the limit 
$\epsilon \rightarrow 0$ ($\tau \rightarrow \infty$) recovers the 
universal limit of spin Hall conductance,
$\sigma_{xy}^z(\omega\rightarrow 0)=e/8\pi$.\cite{sinova04:126603} 
\section{Conclusion}
We calculated the frequency dependent spin susceptibilities for a 2DEG
with both Rashba 
and Dresselhaus spin-orbit coupling.  The suspectibilities have
resonance peaks whose position depends on the relative magnitude of
the Rashba and Dresselhaus coefficients.  The
position of the resonance could be tuned via the gate dependence
of $\alpha$.  We derive a connection between spin density and the
spin current starting from the Heisenberg equation of motion for the
spin density. Unlike the particle density, the resulting continuity
equations have spin current source terms due to the spin-orbit interaction.
Using these relation we connect the  spin current to the
spin density.  In the free electron model the spin current, and the
spin-Hall coefficient, are proportional to the susceptibility.
Indeed, using the continuity equations and the susceptibilities
calculated here we recover the universal (ballistic) value of the
spin-Hall conductance.
Such relations (Eqs.\ (\ref{eq:spinCurrent_x_Exact}) and
(\ref{eq:spinCurrent_y_Exact})) might help elucidate the nature of
spin currents in a similar way to Ref.\ \onlinecite{rashba:0404723}
which discussed 
the relation between the 
spin current and the dielectric function.
By calculating the spin density with the correct impurity
contribution would automatically give the spin current.  
We are confident that these and similar considerations will contribute
to a deeper understanding of the role of impurities in the spin-Hall
effect.
\section{Acknowledgments}
The authors acknowledge financial support from the NCCR Nanoscience,
the Swiss NSF, DARPA, ARO, ONR and the Spintronics RTN.  We would like to
thank J.\ Carlos Egues, Daniel Saraga and Oleg Chalaev for
enlightening discussions. 
%


%
%
\end{document}